\documentclass[%
reprint,
superscriptaddress,
frontmatterverbose,
amsmath,amssymb,
prb,
]{revtex4-2}

\usepackage{graphicx}
\usepackage{subfigure}
\usepackage[normalem]{ulem}
\usepackage{listings}
\usepackage{xcolor}
\usepackage{dcolumn}
\usepackage{bm}

\begin{document}


\title{AD-NEGF: An End-to-End Differentiable
Quantum Transport Simulator}




\author{Zhanghao Zhouyin}
\affiliation{College of Intelligence and Computing, Tianjin University, Tianjin, China.}
\author{Xiang Chen}%
\affiliation{Noah’s Ark Lab, Huawei, Beijing, China.
}%
\author{Peng Zhang}\email{pzhang@tju.edu.cn}
\affiliation{College of Intelligence and Computing, Tianjin University, Tianjin, China.}
\author{Jun Wang}\email{jun.wang@cs.ucl.ac.uk}
\affiliation{%
University College London, London, United Kingdom.
}%
\author{Lei Wang}
\affiliation{Institute of Physics, Chinese Academy of Sciences, Beijing, China}
\date{\today}

\begin{abstract}

The state-of-the-art first principles quantum transport theory and modeling are based on carrying out self-consistent atomistic calculations within the Keldysh nonequilibrium Green's function (NEGF) formalism. The atomistic model of the device can be at the tight-binding (TB) or the density functional theory (DFT) levels, and NEGF determines the nonequilibrium carrier distribution under external bias and gate voltages. In this work, we report an end-to-end automatic differentiable NEGF simulator (AD-NEGF) within the NEGF-TB framework. AD-NEGF calculates gradient information by automatic differentiation (AD) and the implicit layer technique while guaranteeing the correctness of forward simulation. The gradient information enables accurate calculations of transport properties that depend on the derivatives of the transmission coefficient and/or charge current. More interestingly, AD-NEGF can be applied to the extremely interesting inverse design problem, namely, with a desired transport property, AD-NEGF inversely finds a possible device Hamiltonian that would produce such a property.

\end{abstract}

\maketitle




\section{\label{sec_intro}Introduction}


Quantum transport theory provides fundamental understandings of device physics and scientific background knowledge of practical modeling tools for predicting carrier transport in electronic devices \citep{nazarov2009quantum,ryndyk2016theory,wimmer2009quantum}. The state-of-the-art first principles quantum transport theory is based on carrying out atomistic analysis within the Keldysh nonequilibrium Green's function (NEGF) formalism \citep{taylor2001ab, jacoboni2010theory}. Here, the atomic model of the device can be at the tight-binding (TB) level or the density functional theory (DFT) level to capture material details of the device system. The density matrix of the device is constructed by NEGF which provides the nonequilibrium distribution of the carriers under the external bias/gate potentials for the open device structure. A self-consistent NEGF-TB or NEGF-DFT procedure solves the nonequilibrium quantum transport properties including all the atomistic details of the device. Such NEGF methods have been widely applied for device physics and are part of the larger industrial tool-set of technology computer-aided design (TCAD) \citep{silvestri2023hierarchical, medina2021simulation, smidstrup2019quantumatk}.

In practical applications of first principles device modeling, after the self-consistent NEGF simulation is converged, one obtains physical quantities such as the transmission functions $T=T(E)$ where $E$ is the carrier energy, and the electric current $I=I(V)$ where $V$ is the externally applied bias/gate voltages, etc. With $T(E),\ I(V)$, important physical or device parameters that depend on their derivatives, can be further calculated. These include the Seebeck coefficient of thermoelectric devices \citep{kim2011computational}, differential conductance of tunneling spectroscopy \citep{britnell2013resonant}, and the subthreshold swing of MOSFET \cite{prentki2021theory}, etc. To calculate the parametric derivatives of $T(E)$ for instance, a dense energy mesh is usually required especially when $T$ is a rapidly varying function of $E$. In industrial TCAD, one resorts to \emph{compact models} which are analytical models of the carrier transport, in which many measured, fitted and/or phenomenological parameters are used to achieve accuracy. For analytical models, the derivatives can be easily done. For situations where the analytical models do not exist or are difficult to establish, it will be very useful to develop an approach that directly predicts the derivatives of the transport functions without doing brute-force numerical differentiation. In addition, for device physics, being able to predict derivatives or gradients is important in high dimensional optimization of the device models which is related to the inverse problem of property-by-design.

For this purpose, here we report an end-to-end differentiable quantum transport simulator. The automatic differentiable NEGF (hereafter called AD-NEGF) is at the level of NEGF-TB. AD-NEGF calculates gradient information efficiently by automatic differentiation (AD) and implicit layer techniques (see below) while guaranteeing the correctness of forward simulation. The gradient information enables precise calculation of differential physical quantities directly and allows model optimization at a complexity level not achievable by conventional approaches. To the best of our knowledge, we are not aware of end-to-end differentiable quantum transport simulator reported in the literature before.

Our AD-NEGF is inspired by recent progress in AI for quantum transport and differentiable programming. So far, machine learning based AI techniques have been applied to train neural networks with data generated from first-principles transport simulations. Here, the neural network serves as an efficient surrogate model to make predictions of conductance \citep{burkle2021deep,pimachev2021first,li2020neural} and other transport coefficients \citep{lopez2014modeling}. As AI-for-quantum-transport is still in the early stages of development, relatively simple deep learning models such as multi-layer perceptrons \citep{vzupanvcic2020predicting} and convolutional networks \citep{han2021acceleration,souma2021neural,souma2020acceleration} were typically used, although more advanced and specially designed models started to appear \citep{burkle2021deep}. 
Regarding differentiable programming, in our context, it refers to embedding physical models or numerical computation processes into the AI model to improve data efficiency, generalization capability and interpretability. It requires an automatic differentiation framework to support implicit numerical operations such as fixed-point iterations \citep{bai2019deep}, optimization \citep{amos2017optnet}, initial value problems \citep{chen2018neural} etc. Differentiable programming has been applied to physical simulations \citep{hu2019difftaichi,innes2019differentiable} such as rigid body dynamics \citep{de2018end,freeman2021brax}, computational fluid dynamics \citep{kochkov2021machine,holl2019learning,schenck2018spnets}, ray tracing \citep{li2018differentiable} etc. More specifically, in ab-initio simulations, there have been differentiable programming in density functional theory \citep{li2021kohn,kasim2021learning}, Hartree-Fock methods \citep{tamayo2018automatic}, coupled cluster expansions \citep{pavovsevic2020automatic} and molecular dynamics \citep{jaxmd2020}. In the rest of this paper, we present a differentiable programming technique for quantum transport simulations.

We apply AD-NEGF to several situations. First, we demonstrate its ability to accurately and efficiently compute differential physical properties. Second, we demonstrate that combining AD-NEGF with gradient-based optimization can help solve the inverse problem of transport-by-design.
Third, as another inverse problem, we apply AD-NEGF to find possible SKTB parameters of impurity dopants to reach a pre-determined goal of transmission coefficient. We show that AD-NEGF gains significant advantages in these problems over conventional approaches.  The rest of the paper is organized as follows. In the next section, we present details of AD-NEGF. Section III summarizes the applications of AD-NEGF. A short summary is reserved for Section V.

\section{\label{sec:level2}AD-NEGF: theory and implementation}

Our NEGF-TB transport method in AD-NEGF is implemented in PyTorch \citep{NEURIPS2019_9015}. It also includes a Slater-Koster TB (SKTB) module that generates block tri-diagonal TB Hamiltonian \citep{klymenko2021nanonet} of the device material. The back-propagation process in AD-NEGF is improved through the use of implicit gradient techniques, the adjoint sensitivity method for partial differential equations (PDE), and an image charge gradient method.

\subsection{NEGF-TB transport method}

Before discussing the AD process in the next section, it is helpful to briefly present our NEGF implementation on which the AD is applied.
The NEGF first principles quantum transport formalism is based on performing atomistic material specific calculations within the NEGF framework \citep{taylor2001ab}. The atomistic model can be at the level of DFT or TB as mentioned above. This work is based on using TB Hamiltonian for the device material. The idea of the NEGF-TB or NEGF-DFT is to calculate the Hamiltonian of the device under the influence of external bias and gate voltages. In the case of NEGF-TB, the equilibrium Hamiltonian is parameterized by TB parameters and the electrostatic potential due to external electric field is calculated self-consistently. In the case of NEGF-DFT, the entire device Hamiltonian including the effects of external electric field, is calculated self-consistently. On the other hand, the application of NEGF determines the nonequilibrium statistical information for constructing the density matrix. Typically, real space numerical methods are used to handle transport and electrostatic boundary conditions of the open device structure. Since its first report \citep{taylor2001ab}, the NEGF based atomistic modeling methods have become the de facto standard approach for simulating nonequilibrium quantum transport in atomistic nanostructures. For more technical details we refer interested readers to \citet{maassen2012quantum} and in the rest of this section, we outline our implementation on which the AD process is developed. Some further details are summarized in Appendix \ref{sec:details_negf}.


\begin{figure*}
\centering
    \includegraphics[width=1.0\textwidth]{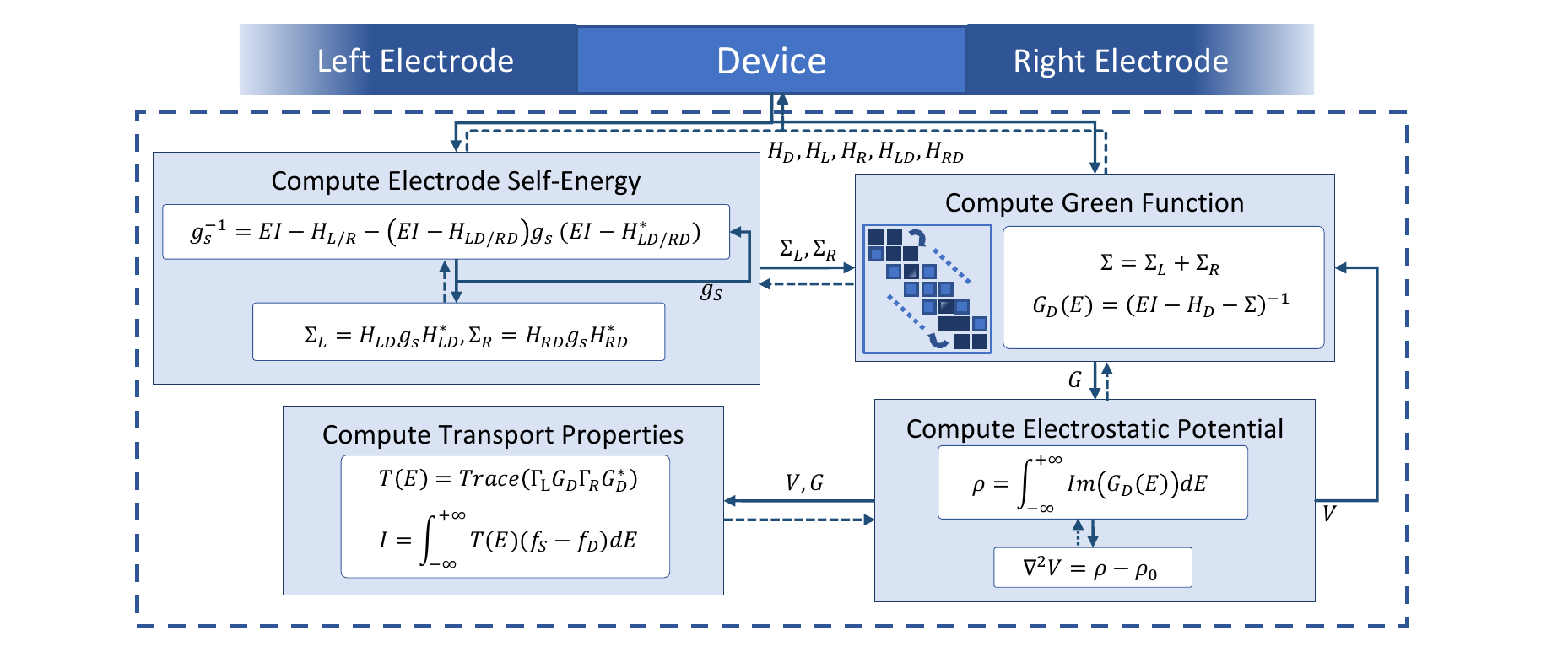}
    \caption{Workflow of AD-NEGF. Solid lines indicate the forward simulation flow, where loops denote self-consistent iterations. Dashed lines indicate the gradient backpropagation flow.}
    \label{fig:negf_flow}
\end{figure*}

Consider a transport system made of a device scattering region and two semi-infinite electrodes that attach to the left and right sides of it, shown in \figurename~\ref{fig:negf_flow}. The Hamiltonian $H$ of the entire system, device and electrodes, are represented with a TB model \citep{slater1954simplified}, which is in a block tri-diagonal form. We assume that a set of orthogonal atomic basis has been applied to reduce the NEGF into this matrix form. The stationary Schr\"{o}dinger equation of the \emph{infinitely large} open device structure is $H\Psi=E\Psi$,
where $\Psi$ is the wave function and $E$ the corresponding energy. The Green's function of the system is formally obtained as:
\begin{equation}
    G=[EI-H]^{-1},
    \label{eq1}
\end{equation}
where $I$ is the identity matrix. Note that for the open device structure, the Hamiltonian $H$ is in fact an infinitely large matrix thus the Green's function $G$ cannot be directly obtained from Eq.~(\ref{eq1}). This problem is resolved by computing the Green's function only for the device scattering region $G_D$ via its Hamiltonian $H_D$, and the electronic degrees of freedom in the two semi-infinite electrodes are integrated out, resulting in the self-energy $\Sigma$ terms which are added to $H_D$. $G_D$ is solvable since it is finite. The Green's function $G_D$ and its conjugate construct the Keldysh NEGF $G^<$ via the Keldysh equation will give the non-equilibrium charge distribution in the device \citep{maassen2012quantum}. With the charge distribution, Poisson's equation for the electrostatic potential in the device scattering region $V_D$ is solved which updates the Hamiltonian $H_D$. This process is self-consistently iterated to numerical convergence.
Following this standard procedure \citep{taylor2001ab,maassen2012quantum}, $G_D$ is obtained by inverting a finite matrix,
\begin{align}
    G_{D}=[EI-H_{D}-\Sigma]^{-1}.
    \label{device_green}
\end{align}
Since the matrix to be inverted can be cast into block tri-diagonal form due to the short range-ness of the TB potential,  we apply the efficient recursive algorithm in \citep{anantram2008modeling} to obtain $G_D$. In Eq.~(\ref{device_green}), the self-energy $\Sigma$ due to the two electrodes of the device can be calculated using the surface Green's function technique \citep{sancho1985highly}, the details of which are summarized in Appendix \ref{self-ener}.

An electrostatic potential $V_D$ in the device scattering region is established due to external bias/gate voltages applied to the device.  $V_D$ is solved self-consistently via Poisson's equation and added to the $H_D$,
\begin{align}
\begin{cases}
    \nabla\cdot\epsilon(r)\nabla [\Delta V_D(r)]&=-[\rho(r;\Delta V_D)-\rho_0(r)], \\
    \Delta V_D(r)|_{\{z_L,z_R\}}&=\{V_L, V_R\},
    \label{Poisson}
\end{cases}
\end{align}
where $V_L$ and $V_R$ are boundary conditions at electrodes $z_L$ and $z_R$, $\Delta V_D=V_D-V_0$ is the difference between the real potential and the equilibrium one. The charge density $\rho$ on the right-hand side is obtained by NEGF, details summarized in Appendix \ref{appendix-poisson}. In our Poisson's equation solver, an efficient image charge approach based on the fast multipole method (FMM) \citep{svizhenko2005effect,zahn1976point} is used. After $V_D$ is obtained, it is added to $H_D$ to calculate Green's functions and the process is repeated until self-consistency.

Once the $G_D$-$V_D$ self-consistency reaches a small numerical tolerance, we calculate the transmission coefficient $T(E)$ by Landauer formula and further the I-V curves $I(V)$ by integrating the $T(E)$ over the bias window. Details of the implementation are summarized in Appendix \ref{appendix-transport-expressions}.

\subsection{Differentiable NEGF}

Our differentiable NEGF model is implemented in PyTorch \citep{NEURIPS2019_9015}. When executing a program, PyTorch will automatically track the functions to build a computational graph by their calling orders, so that after the entire program is executed, the corresponding gradient can be computed by running backward through the computational graph based on the chain rule. However, for some numerical processes, their implementation is either unavailable in the PyTorch framework, (e.g. Poisson's equation solver), or includes iterative processes such that the computation graphs are too large to track (e.g. self-consistent iteration). For this purpose, we extend the PyTorch gradient computation with implicit gradient techniques for backpropagation through self-consistent iterations and using an adjoint sensitivity method for calculating gradients through the Poisson's equation \citep{pontryagin1987mathematical}. In addition, an efficient gradient formula for the image charge \citep{svizhenko2005effect} - accelerated by FMM, is developed to accelerate the gradient calculation. This formulation can be regarded as a summation of point charges produced by the gradients which can also be computed with FMM.

Regarding the implicit gradient technique, it is needed when direct automatic differentiation through function $y=f(x)$ is unavailable or expensive to compute. Instances often arise when one wishes to calculate gradients through numerical solvers or complicated iterative algorithms. Based on the implicit function theorem \citep{krantz2002implicit}, if there exists such constrained function $h(y,x)=0$ where $y$ is taken as the converged output of function $f$, the gradient $\frac{dy}{dx}$ is obtained as:
\begin{align}
\frac{dy}{dx}=-\left[\frac{\partial{h(y,x)}}{\partial{y}}\right]^{-1}\frac{\partial{h(y,x)}}{\partial{x}}.
\label{eq:implicit}
\end{align}
We use the implicit gradient to derive the gradient of the surface Green's function \citep{sancho1985highly} that appears in the self-energy $\Sigma$ calculation (see Appendix \ref{self-ener}).  In particular, the converged surface Green's function $g_s(\theta)$ in Appendix \ref{self-ener} must satisfy the self-consistent equation (\ref{sgf}). Hence $h(g_s, \theta)=[A_{ll}-A_{ll-1}g_sA_{l-1l}^\dagger]-gs^{-1}=0$, where $A_{ll}$ stands for $[ES_{ll}-H_{ll}]$, and $\theta$ denotes the input variables to compute $g_s$. Thus we can write down the gradient of $g_s$ with respect to $\theta$ explicitly by
\begin{align}
\frac{dg_s}{d\theta}=-\left[\frac{\partial{h(g_s,\theta)}}{\partial{g_s}}\right]^{-1}\frac{\partial{h(g_s,\theta)}}{\partial{\theta}}.
\label{eq:implicit_sg}
\end{align}

Another place that the implicit gradient is applied, is to compute gradients through the self-consistent Poisson's equation under external electrostatic boundary conditions. Note that Poisson's equation \ref{Poisson} depends on charge density $\rho$, while the charge density is given by the density matrix via NEGF in Eq.~(\ref{A10}), also shown in the self-consistent loop of \figurename~\ref{fig:negf_flow}. To perform back-propagation through Poisson's equation solver, adjoint sensitivity method \citep{plessix2006review,pontryagin1987mathematical} for PDE-constrained optimization is adopted which is a technique for constrained optimization in inverse problems. Here, the forward process of the numerical PDE solver is unaltered which is often denoted as the state equation that links the controlled parameter and the state of the constrained system. Meanwhile, an adjoint state equation that connects the perturbation of variables and states is solved by using the same numerical solver. The gradients can then be evaluated with the adjoint state, and join in the gradient chain of backward propagation. Since the adjoint state equation is often independent of the number of controlled variables, the total complexity is proportional to the forward process which makes it suitable for control problems with scalar output and high dimensional inputs. Recently, the adjoint method has been applied in constructing subtle neural networks containing dynamic physical processes including neural ODE \citep{chen2018neural} and deep equilibrium model \citep{bai2019deep}, which can be considered as examples of cooperations of auto-differentiation and adjoint methods. For our problem here, since in TB models
the electrostatic potential is established by point charges, $\Delta q(r)=\sum_i\Delta q_i\delta(r-r_i)$, we developed a method to evaluate the gradients of such situations. By linearity of the Poisson's equation, the original form is decomposed into a Laplace's equation with Dirichlet boundary condition and a Poisson's equation with zero Dirichlet boundary condition,
\begin{align}
&\begin{cases}
    -\nabla^2(\Delta V_1(r))=0, \\
    \Delta V_1(r)|_{\{z_L,z_R\}}=\{V_L,V_R\}.
\end{cases}
\\
&\begin{cases}
    -\nabla^2(\Delta V_2(r))=\frac{1}{\epsilon}\Delta \rho(r),\\
    \Delta V_2(r)|_{\Sigma}=0.
\end{cases}
\end{align}
Laplace's equation can be easily solved. The second equation can be solved with image charges \citep{svizhenko2005effect,harb2019scattering}, and the second potential can be written as:
\begin{align}
    V_2(r_i)&=\sum_{j\in{N},j\neq i}\frac{q_j}{4\pi\epsilon}\frac{1}{\sqrt{t_{ij}^2+(z_i-z_j)^2}} \nonumber \\ &+\sum_{j\in{N}}\frac{q_j}{4\pi\epsilon}\sum_{n=1}^\infty\left[\frac{1}{\sqrt{t_{ij}^2+\Delta_1^2}}-\frac{1}{\sqrt{t_{ij}^2+\Delta_2^2}}\right.\nonumber\\&\left.+\frac{1}{\sqrt{t_{ij}^2+\Delta_3^2}}-\frac{1}{\sqrt{t_{ij}^2+\Delta_4^2}}\right],
\end{align}
where $t_{ij}^2=(x_i-x_j)^2+(y_i-y_j)^2$, and $\Delta^2$ stands for the distance in the transport direction between central charges and charges from two electrodes. Therefore, the first term here describes the interactions inside the device, while all the remaining terms simulate the effect of its coupling to charges outside the device scattering region. The summation of the second term is computed until achieving high accuracy which,  empirically, requires hundreds of sites. To speed up this calculation, we apply the FMM \citep{engheta1992fast} to reduce the computational complexity from $O(N^3)$ to $O(N^{4/3})$, where $N$ is the number of sites. To perform backward propagation through the fast multipole layer, the gradient of the output potential to the charges is required. By taking the derivative of a target objective $L: C^d \xrightarrow{} R$, the derivative of $L$ with respect to charge $q_j$ can be expanded as the image summation form of accumulated gradients from the last layer, which is:
\begin{align}
    \label{eq:charge_gradient}
    \frac{\partial L(V)}{\partial q_j} = & \sum_i\frac{\partial L}{\partial V_i}\frac{\partial V_i}{\partial q_j} \nonumber\\
    = & \sum_{i\in{N},i\neq j}\frac{\partial L/\partial V_i}{4\pi\epsilon}\frac{1}{\sqrt{t_{ij}^2+(z_j-z_i)^2}} \nonumber \\
    & + \sum_{i\in{N}}\frac{\partial L/\partial V_i}{4\pi\epsilon}\sum_{n=1}^\infty\left[\frac{1}{\sqrt{t_{ij}^2+\Delta_1^2}}-\frac{1}{\sqrt{t_{ij}^2+\Delta_2^2}}\right.\nonumber\\&\left.+\frac{1}{\sqrt{t_{ij}^2+\Delta_3^2}}-\frac{1}{\sqrt{t_{ij}^2+\Delta_4^2}}\right].
\end{align}
Similarly, computing gradients of this form can be accelerated by FMM with a complexity of $O(N^{4/3})$, which is significantly faster than iteratively solving the adjoint Poisson's equation.

In summary, AD-NEGF is realized by the following steps:
\begin{enumerate}
    \item The entire calculation is implemented in PyTorch so that the explicit numerical expressions are automatically differentiated by PyTorch.
    \item For the implicit equations such as the self-consistent iterations for the surface Green's function and the non-equilibrium charge densities, as well as Poisson's equation, we implement the corresponding numerical solvers in the PyTorch autograd forward functions, and implement the gradient computation methods in the corresponding PyTorch autograd backward functions (i.e., implicit gradient of Eq.~(\ref{eq:implicit}) for iterative solvers and charge gradient of Eq.~(\ref{eq:charge_gradient}) for Poisson's equation solver). Therefore, in such cases, the gradients through the numerical solvers are computed by our customized algorithms instead of automatic differentiation.
    \item By implementing the above steps, the gradient of the entire NEGF-TB process can be computed end-to-end simply by backpropagation.
\end{enumerate}

\begin{figure*}[tb]
    \centering
    \subfigure[Structure of an AGNR with width 7 and length 5.]{
    \includegraphics[width=0.3\linewidth]{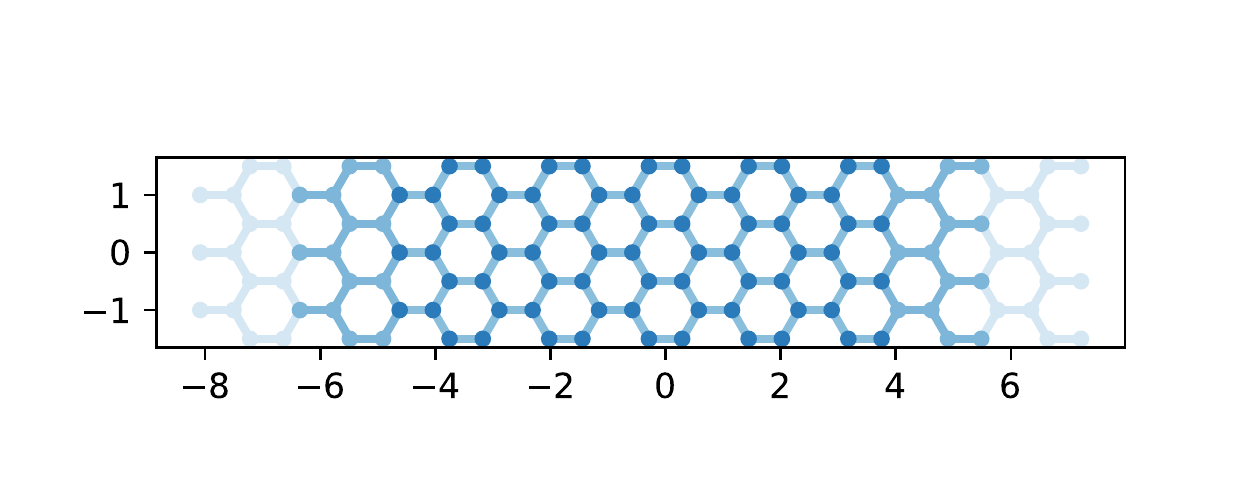}
    \label{fig:AGNR}}
    \subfigure[Structure of a 7-4 graphene nano-junction.]{
    \includegraphics[width=0.3\linewidth]{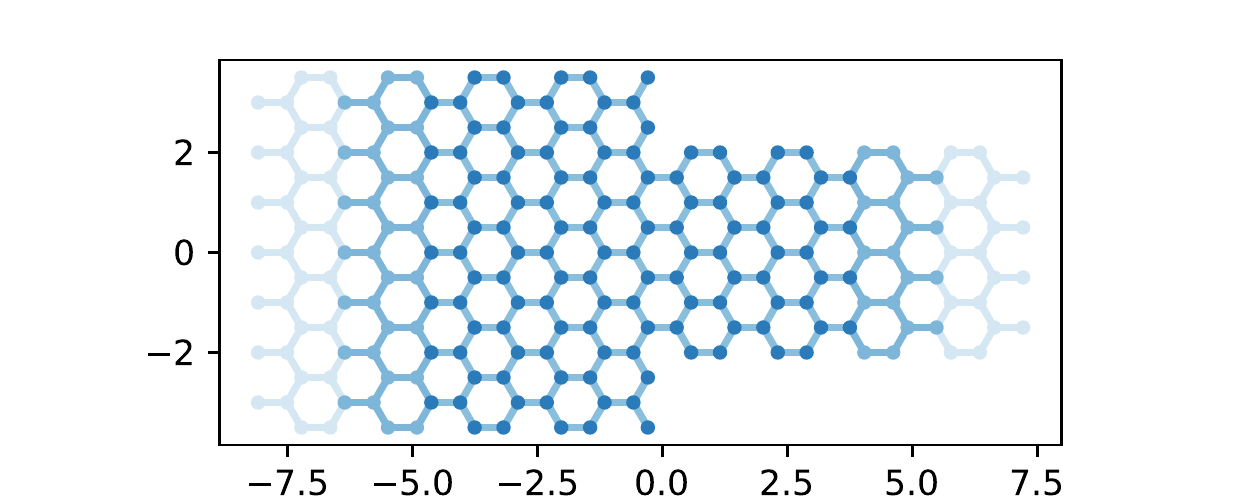}
    \label{fig:7_4_nano_junction}}
    \subfigure[Structure of a 5-2 graphene nano-junction.]{
    \includegraphics[width=0.3\linewidth]{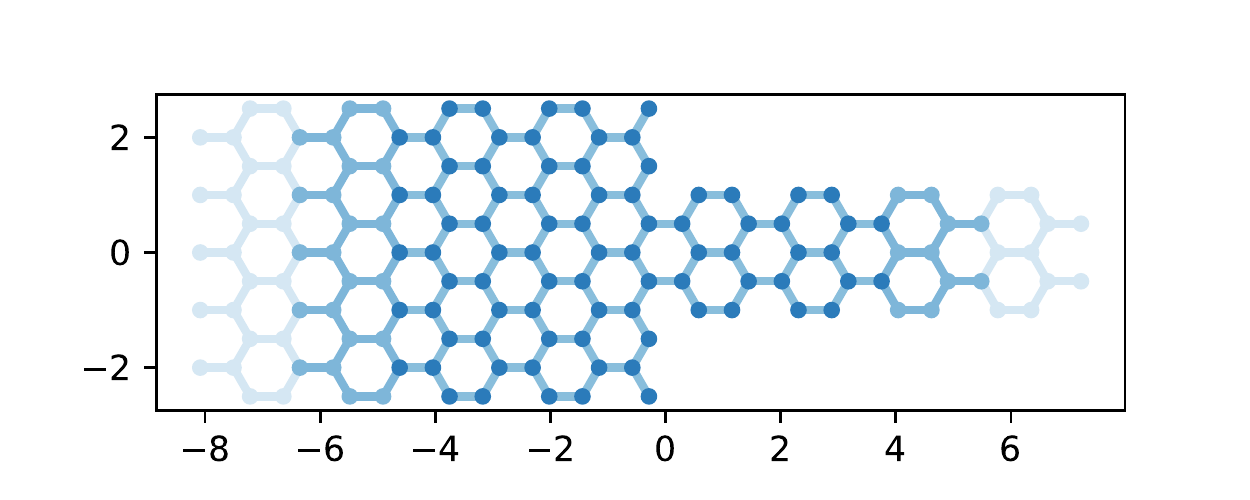}
    \label{fig:5_2_nano_junction}}
    \caption{Device structures used in the experiments.}
    \label{fig:structures}
\end{figure*}

\section{Examples of AD-NEGF}

We have applied AD-NEGF to two-probe transport junctions made of armchair graphene nanoribbon (AGNR), shown in \figurename~\ref{fig:structures}. More details of the calculation parameters can be found in Appendix \ref{appendix:repro}.

\begin{figure}[tb]
    \centering
    \subfigure[Transmission and DOS calculated by AD-NEGF and confirmed with ASE.]{
    \includegraphics[width=0.9\linewidth]{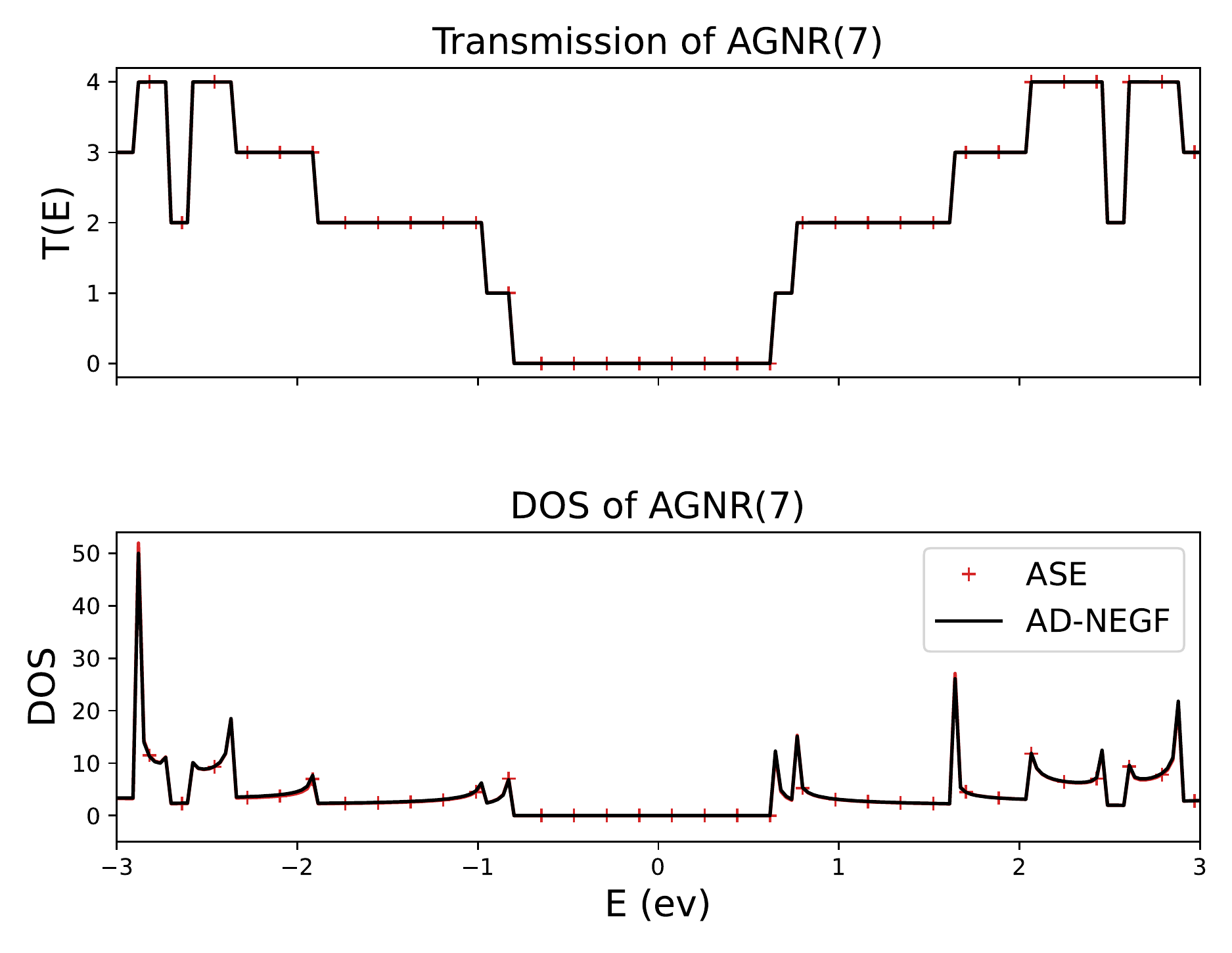}
    \label{fig:quantities_a}}
    \\
    \subfigure[Seebeck coefficient and differential conductance calculated by AD-NEGF.]{
    \includegraphics[width=1.0\linewidth]{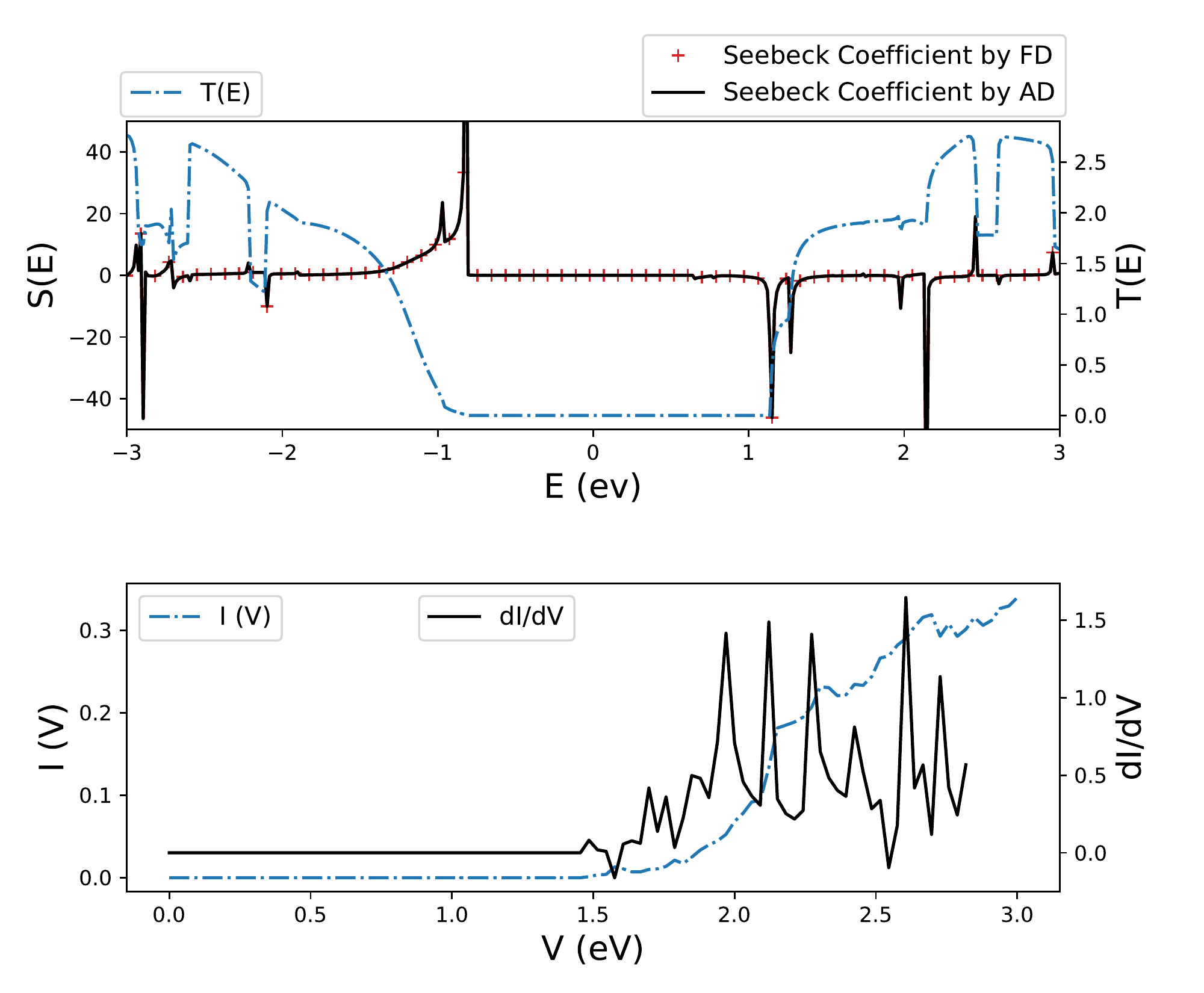}
    \label{fig:quantities_b}}
    \caption{Transmission quantity computation with AD-NEGF.}
\label{fig:quant}
\end{figure}

\subsection{Differential Transmission}

Differential transmission is needed when calculating physical quantities such as the Seebeck coefficient and differential conductance. The Seebeck coefficient, also known as thermoelectric power, measures the induced voltage across a transport junction in response to a temperature gradient. Theoretically, the Seebeck coefficient is related to the derivative of the transmission function $T(E)$ versus the energy $E$, evaluated at the chemical potential of the system \citep{reddy2007thermoelectricity}: 
\begin{align}
S=-\frac{\pi^2k_{B}^2\Upsilon}{3e}\frac{\partial{ln(T(E))}}{\partial{E}},
\label{Seebeck1}
\end{align}
where $\Upsilon$ is the temperature and $k_B$ the Boltzmann constant. The differential conductance is another very useful quantity that is related to differential transmission. It is commonly used to analyze nonlinear current-voltage characteristics in tunneling spectroscopy, and devices with negative differential conductance are used in electronic oscillators and amplifiers. Theoretically, the differential conductance is obtained by the gradient of electric current to applied bias voltage: $G = \frac{d{I}}{d{V}}$.

For the AGNR system of width 7 and length 5, as shown in \figurename~\ref{fig:AGNR}, the transmission function $T(E)$ and the density of states (DOS) are calculated by our AD-NEGF, shown in \figurename~\ref{fig:quantities_a}. The results are in perfect agreement with those obtained by ASE \citep{larsen2017atomic}. AD-NEGF is then deployed to obtain  the Seebeck coefficient by Eq.~(\ref{Seebeck1}) and the differential conductance, results shown in \figurename~\ref{fig:quantities_b}. The step-like transmission function $T(E)$ leads to singular behavior in its derivative, giving rise to peaks in the Seebeck coefficient curve. While a direct brute-force calculation of the differentiation can be done (FD, red pluses in \figurename~\ref{fig:quantities_b}), such calculation is highly sensitive to the fine energy mesh. In comparison, AD-NEGF gives precise values of the differentiation at any energy point (black curve). In particular, for direct numerical differentiation, the trade-off between the truncation error and the round-off error is observed by selecting different energy mesh sizes from $10^{-2}$ to $10^{-5}$ eV. With a coarse mesh, peaks in the Seebeck coefficient may be missing or mistakenly generated due to the truncation error. With a very fine mesh, lacking machine precision causes significant noise which may lead to meaningless results. In addition to accuracy, evaluating the Seebeck coefficient with AD-NEGF is also faster than numerical differentiation by roughly 30\%, due to our particular back-propagation procedure in AD-NEGF.

To summarize, the correctness and effectiveness of AD-NEGF are validated in comparison with direct brute-force numerical differentiation. In AD-NEGF, differential transport quantities are calculated by simply calling a single backward step. Moreover, the process of computing derivatives is itself differentiable, permitting the computation of higher-order derivatives such as the nonlinear conductance coefficients \citep{ma1999weakly}.


\begin{figure}
    \centering
    \includegraphics[width=1.0\linewidth]{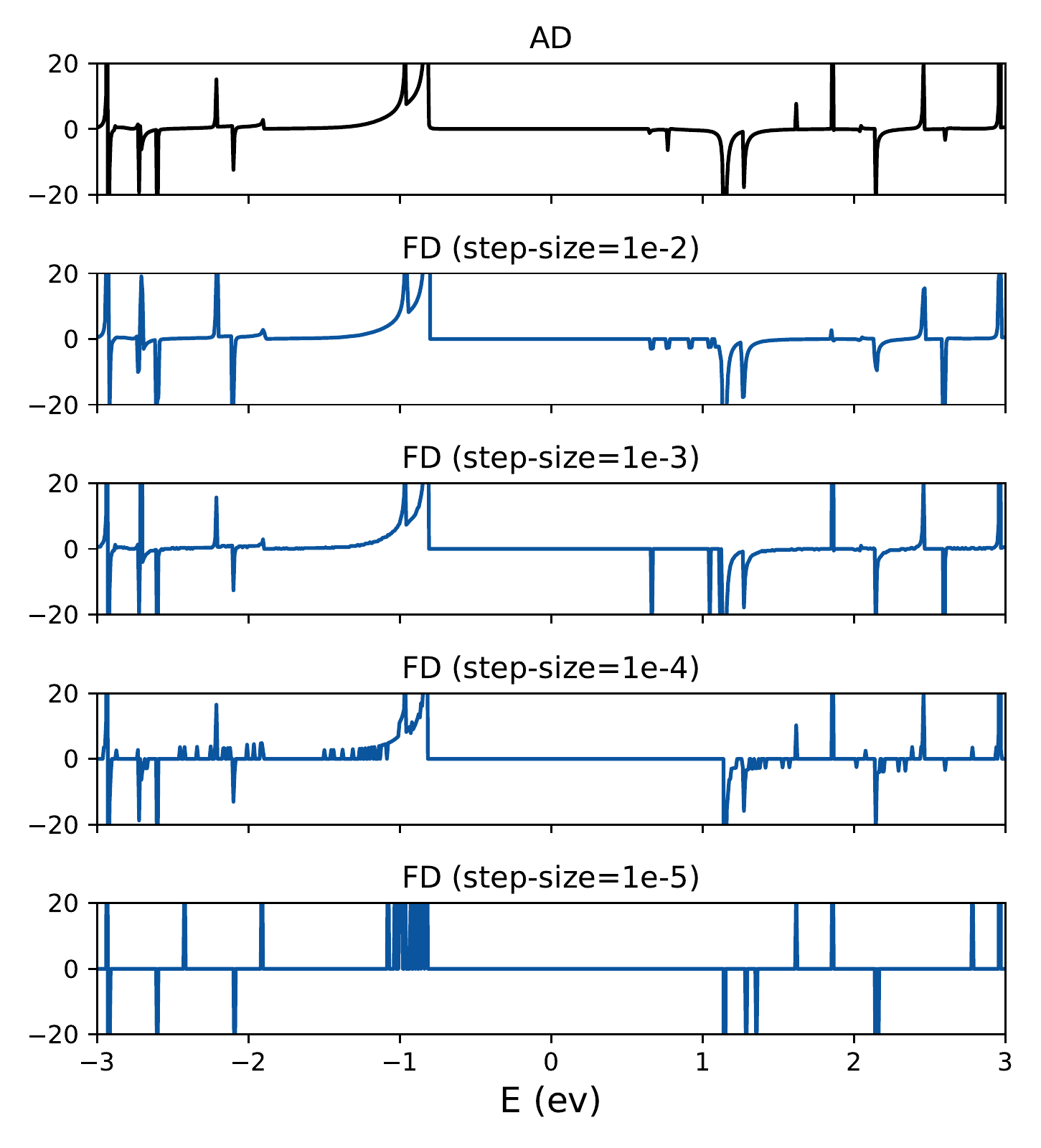}
    \caption{Comparison of automatic differentiation and numerical differentiation with different step-sizes.}
    \label{fig:seebeck}
\end{figure}

\begin{figure}
    \centering
    \includegraphics[width=1.0\linewidth]{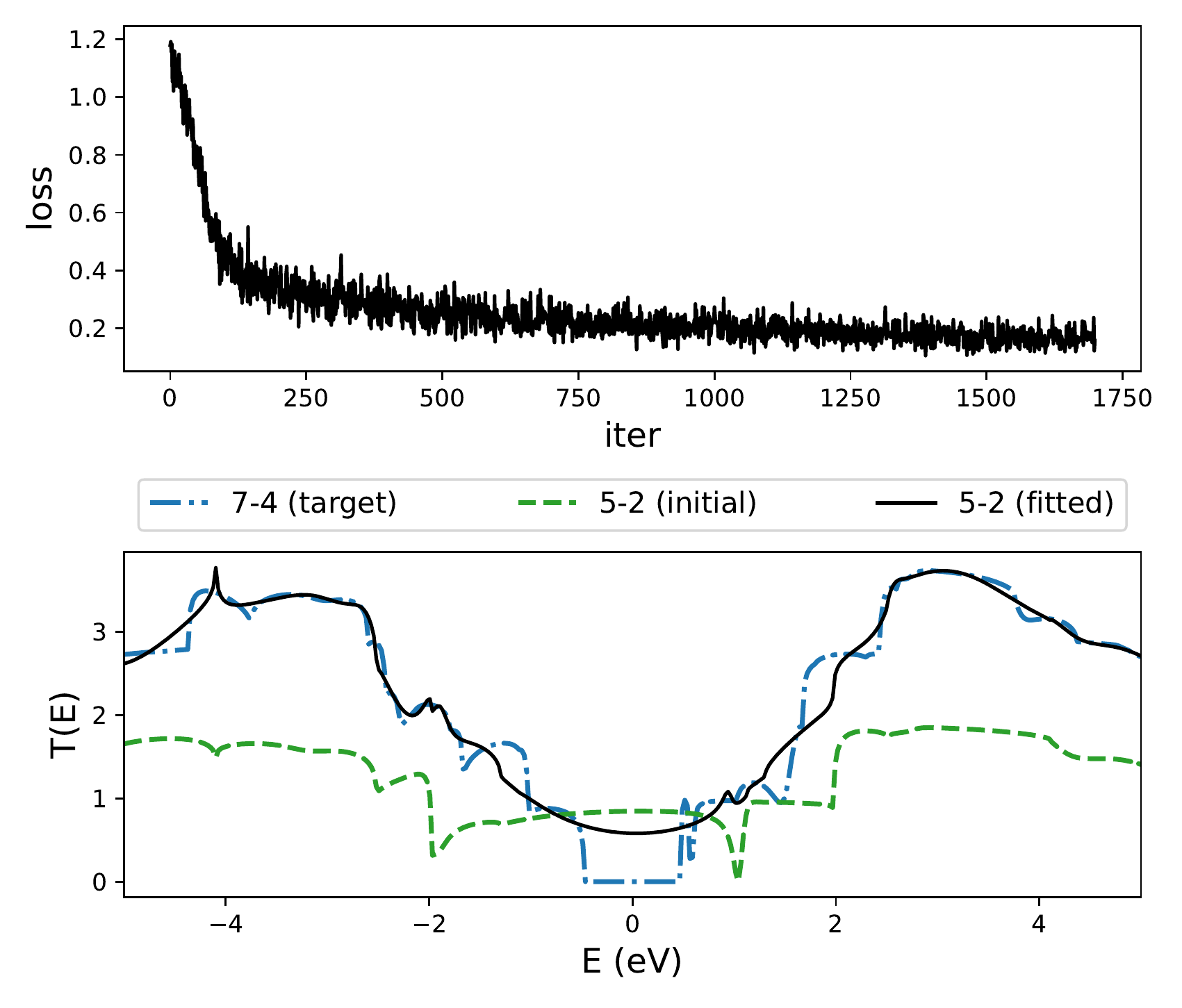}
    \caption{The fitting loss and the fitted transmission curve of a 5-2 graphene nano-junction.}
    \label{fig:fitting}
\end{figure}


\subsection{Transmission by design}


In this section, we show that AD-NEGF can be potentially useful to give insight to the problem of transport-by-design. Namely, if one wishes to obtain a desired transport property, can one design a Hamiltonian that does produce it?  Such an inverse problem is very difficult - if not impossible to solve, here we show AD-NEGF may lead to a possible route. In general, the inverse problem is about inferring input parameters reversely from the objective. One approach is using black-box optimization methods to sample a large number of input combinations, but the computational cost grows exponentially with the number of parameters, making it intractable for this task. On the other hand, based on AD-NEGF, the gradient of the transport property with respect to the Hamiltonian elements can be computed by simply calling the forward and backward computation each for one time, the computational complexity of which is irrelevant of the number of parameters to be determined in the Hamiltonian. Such characteristics of AD-NEGF allow for conducting gradient-based optimization on the Hamiltonian elements to fit the desired properties. Through such paradigms, AD-NEGF holds the potential to solve transport-by-design problems in material science.

Let's consider a 7-4 graphene nano-junction, consisting of 7 graphene rings on the left and 4 rings on the right. The transmission coefficient $T_{74}(E)$ of this system is calculated by NEGF-TB, shown as the blue dash-dotted curve in the lower panel of \figurename~\ref{fig:fitting}, which serves as the desired result. In this toy exercise of the inverse problem, we wish to find a Hamiltonian that will produce this $T_{74}(E)$. To this end, we may start from any physically sound Hamiltonian as an initial guess, for example, the Hamiltonian of a 5-2 graphene nano-junction $H_{52}$, which produces a totally different transmission $T_{52}(E)$ as depicted by the green dashed curve in \figurename~\ref{fig:fitting}. With the $T_{74}(E)$ as the goal and using the gradient-based optimization in AD-NEGF, it is possible to automatically vary the parameters in $H_{52}$ such that it generates the desired result $T_{74}$. The fitting parameters are the elements of $H_{52}$ including the device, leads, and the corresponding couplings. For this exercise, the dimension of the optimizing variables is at the level of $10^4$. The transmission curve, as shown in \figurename~\ref{fig:fitting}, consists of 2000 energy points sampled from (-5eV, 5eV). Since directly computing the gradients of all 2000 points is inefficient, we apply the stochastic gradient descent algorithm to conduct mini-batch optimization which has shown supremacy of efficiency and performance in high dimensional optimization problems. The fitting parameters are optimized with the Adam optimizer \citep{kingma2014adam} built in PyTorch, making the procedure highly similar to training a neural network.

The results are displayed in \figurename~\ref{fig:fitting}. The upper panel shows the loss function versus the optimization iteration, where the loss is reduced to a considerably low level after a few hundred iterations, which means the converged $H_{52}$ parameters of the 5-2 nano-junction could approximately produce $T_{74}(E)$ of the 7-4 nano-junction. Indeed, the black solid curve in the lower panel, obtained by the converged $H_{52}$, is akin to a smoothed $T_{74}$ of the 7-4 junction. This is consistent with the intuition since a graphene junction of 5-2 has less degree of freedom than that of a 7-4 nano-junction. We mention in passing that we have also tried traditional black-box optimization methods including Bayesian optimization, genetic algorithm, and gradient-based optimization with numerical differentiation, but none works for this problem because of the curse of dimensionality. 

Finally, we wish to mention that solving the problem of transport-by-design can be potentially very useful in applications where a particular transport property is desired. As we have shown here, AD-NEGF can inversely determine a Hamiltonian that would approximately generate the desired property. Since the Hamiltonian matrix elements are made of atomic potentials, it would provide tremendous intuitions on the material and external manipulation  (i.e. stress, doping, impurity, external fields etc) to produce the desired transport.

\subsection{On-site doping}
\begin{figure*}
    \centering
    \subfigure[Loss against running time and iteration steps respectively.]{
    \includegraphics[width=0.47\textwidth]{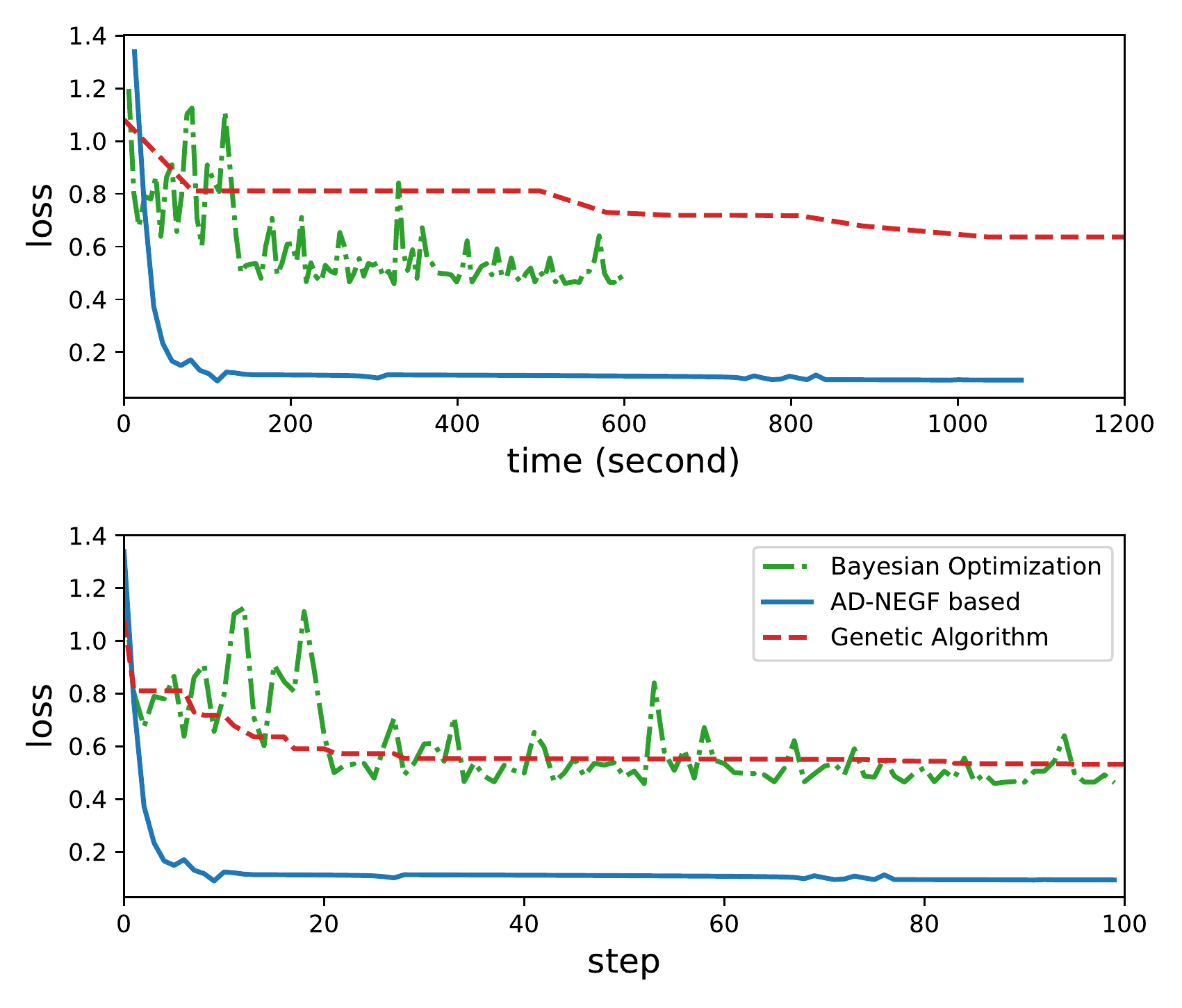}
    }
    \hspace{-6mm}
    \subfigure[Original and optimized transmission curves.]{
    \includegraphics[width=0.51\textwidth]{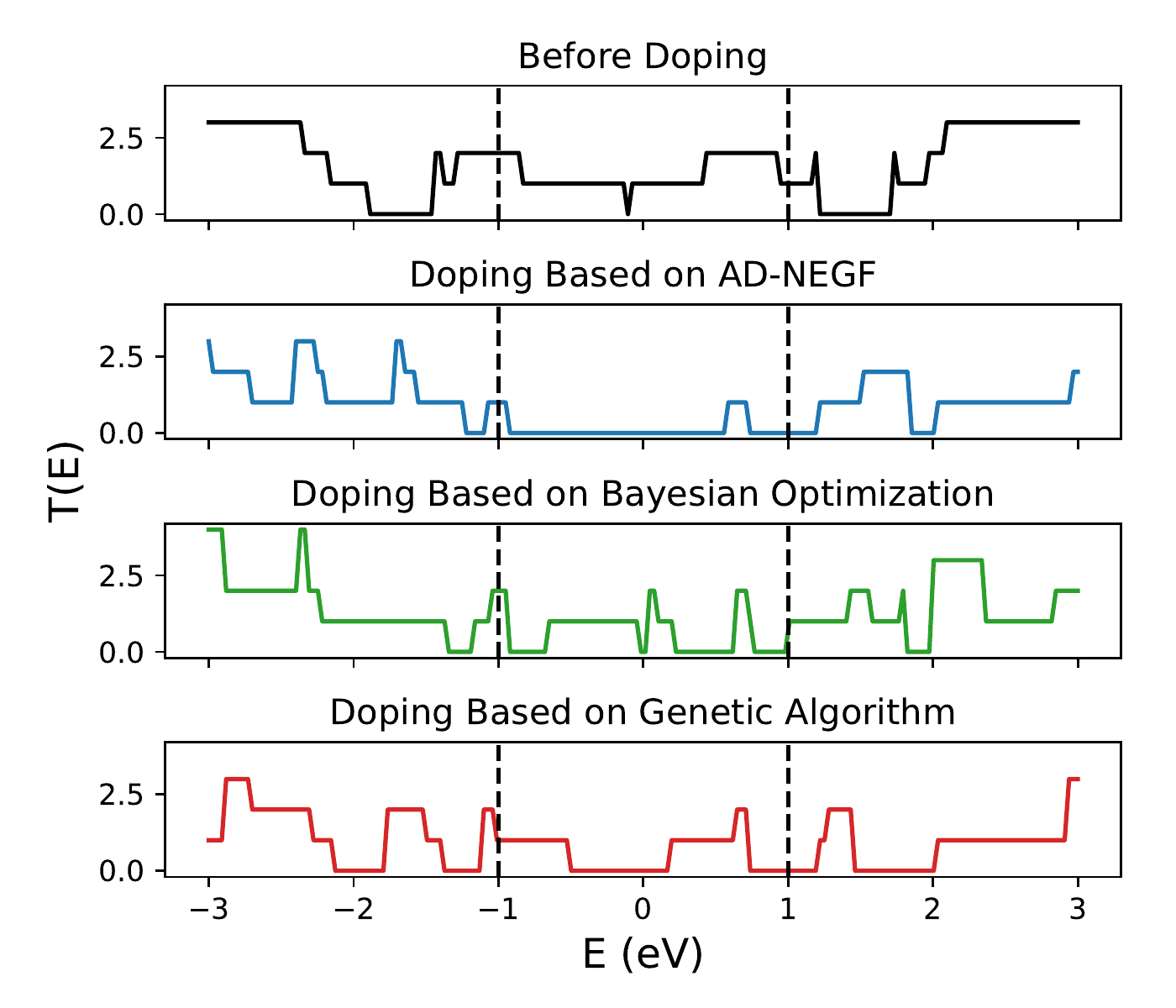}
    }
    \caption{Comparison between AD-NEGF and conventional black-box optimization methods in the doping optimization task.}
    \label{fig:dop}
\end{figure*}

\begin{figure}
    \centering
    \includegraphics[width=1.0\linewidth]{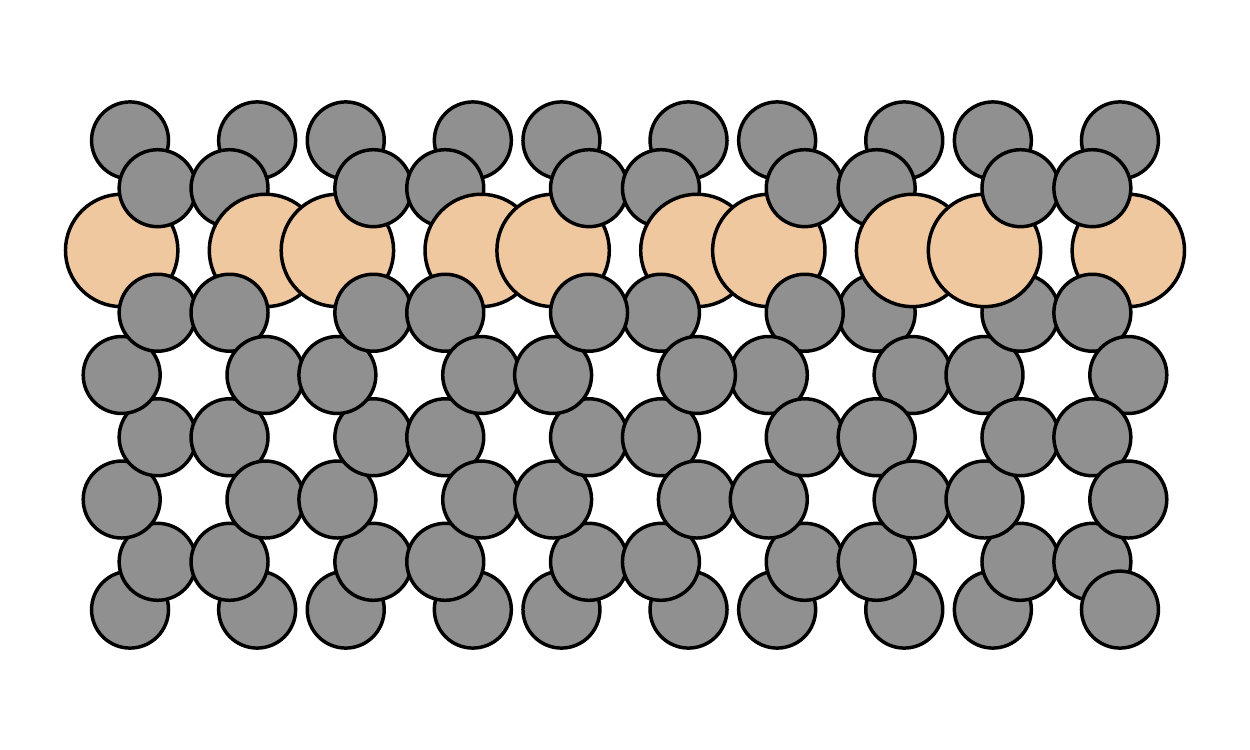}
    \caption{The structure of the doped AGNR system, where the atoms depicted in Navajo white are replaced with impurities.}
    \label{fig:dop_stru}
\end{figure}

Modern device engineering is capable of manipulating material properties at the atomic level. By stress and impurity doping etc., electronic structure and material parameters can be controlled and modified for better device performance. Here we further explore the possibility to solve practical inverse problems with AD-NEGF through an end-to-end doping optimization. The doped structure is illustrated in \figurename~\ref{fig:dop_stru}, where some of the carbon atoms (in grey) are replaced with the impurities (in Navajo white). To this end, as the desired goal we wish to reduce the average transmission of the AGNR in a specified energy range of (-1eV, 1eV), by doping impurity atoms into the scattering region.
Doping can be modelled as an effective change in the site and the hopping parameters in the TB Hamiltonian, i.e., the diagonal and off-diagonal elements of the Hamiltonian matrix. In contrast to the inverse problem presented in the last section, here we are only allowed to vary the SKTB parameters associated with the dopant atoms, leaving parameters of the host material not touched. We may view this as an optimization problem of the SKTB parameters associated with the dopants to reach the desired goal, and the SKTB parameters include orbital energy and two-center integrals. The total number of optimization variables is 13 since only one impurity atom is replacing a carbon atom in the AGNR.

For comparison, we also apply conventional optimization methods of genetic algorithm and Bayesian optimization. The results are displayed in \figurename~\ref{fig:dop}. In the loss diagram on the left-panels, the gradient-based method in AD-NEGF converges significantly faster and much better than the conventional approaches in terms of computational time as well as the total iteration steps. We also found that the conventional optimizations are sensitive to the preset hyper-parameters, and were not able to reach the loss-level of AD-NEGF (see left panels of \figurename~\ref{fig:dop}). Corresponding to the loss curves, the AD-NEGF has essentially reached our design goal of reducing 
transmission in the energy range of (-1eV, 1eV), shown in the right panels of \figurename~\ref{fig:dop} (blue curve). In comparison, the conventional optimization was not able to reach the design goal (green and red curves). 

These results validate the effectiveness of the AD-NEGF method in conducting practical atomic-level inverse design to optimize transport properties by cooperating with material models.

\section{Summary}

Motivated by recent advances in AI for quantum transport and differentiable programming, we have developed an automatic differentiation capability into the atomistic NEGF quantum transport simulator. The end-to-end automatic differentiable NEGF simulator, AD-NEGF, calculates gradient information by AD while guaranteeing the correctness of forward simulation. The gradient information enables accurate predictions of transport properties that depend on the derivatives of the transmission coefficient and/or charge current, such as the Seebeck coefficients in the thermoelectric phenomenon and differential conductance in nonlinear carrier transport. For ballistic transport in confined nanostructures, transmission functions often vary rapidly as a function of carrier energy due to quantum interference, causing its derivative to be singular thus hard to accurately determine. AD-NEGF solves such problems very accurately as we demonstrated in this work. In the experimental technique of inelastic tunneling spectroscopy (IETS) \citep{reed2008inelastic}, the opening of phonon-assisted transmission channels leads to slight changes of the measured current at a certain bias voltage, and the signal is picked up by measuring the differential conductance. In NEGF simulations of IETS, AD-NEGF can be used to directly compare with the measured signal. AD-NEGF can also be very useful for accurate and efficient simulations of other transport coefficients related to nonlinear expansions of current and/or charge versus external voltages.

More interestingly, AD-NEGF can be applied to the inverse design problem, namely, with a desired transport property, AD-NEGF inversely finds a possible device Hamiltonian that would produce such a property. While property-by-design is a dream goal of materials and device physics, it is an extremely difficult problem to solve. In particular, due to the high dimensionality of Hamiltonian matrices in NEGF simulations, conventional optimization techniques are essentially powerless for such inverse problems. To this end, we showed that AD-NEGF with gradient-based optimization has great potential. Here we showed that starting from a pre-defined transport property (transmission function), AD-NEGF inversely determines a possible Hamiltonian that would approximately produce it. Though the examples are relatively simple, the idea is very clear. In real practical applications, once the model Hamiltonian is inversely determined, one may investigate its on-site and hopping parameters which would generate deep insights in how to realize such a model with real materials. 

\begin{acknowledgments}
The authors would like to thank Prof. Hong Guo for useful discussions on quantum transport theory and for his critical reading of the manuscript before submission. This work is supported in part by the Natural Science Foundation of China (grant No.62276188). L.W. is supported by the Strategic Priority Research Program of the Chinese Academy of Sciences under Grant No. XDB30000000 and National Natural Science Foundation of China under Grant No. T2121001.
\end{acknowledgments}

\appendix
\section{Additional Details on the NEGF Method}
\label{sec:details_negf}

\subsection{Self-Energy}
\label{self-ener}
The self-energy of electrodes is computed from the surface green function $g_s$ of the electrode layer coupled with devices. Here we assume that the system is made up of a device and two semi-infinite contacts on the side. Equation (\ref{eq1}) can be expanded in the following form:
\begin{align}
\begin{bmatrix}
A_{L}&A_{LD}&0\\
A_{DL}&A_{D}&A_{DR}\\
0&A_{RD}&A_{R}
\end{bmatrix}
\begin{bmatrix}
G_{L}&G_{LD}&G_{LR}\\
G_{DL}&G_{D}&G_{DR}\\
G_{RL}&G_{RD}&G_{R}
\end{bmatrix}
=I,
\end{align}
where $A=[EI-H]$, and the subscripts are used to distinguish the matrix elements corresponding to the left lead (L), the device (D), the right lead (R), and their interactions. Thanks to its block tri-diagonal form, the device Green function $G_{D}$ satisfies:
\begin{align}
[A_{D}-A_{DL}A_{L}^{-1}A_{LD}-A_{DR}A_{R}^{-1}A_{RD}]G_{D}=I.
\end{align}
Since $A_D=[EI-H_D]$, compared with Equation (\ref{device_green}), we have
\begin{align}
    \Sigma^L &= A_{DL}A_{L}^{-1}A_{LD},\\
    \Sigma^R &= A_{DR}A_{R}^{-1}A_{RD},\\
    \Sigma &= \Sigma^L + \Sigma^R.
\end{align}
To avoid using full leads Hamiltonian, it is assumed that only the neighboring layers have interactions with each other. We denote the left lead layer connected to the device by $l$. Then the left self-energy can be simplified as $\Sigma^L=A_{Dl}A_{l}^{-1}A_{lD}$. The coupling matrix $A_{lD}$ is given as input of NEGF. What remains unclear is $A_{l}^{-1}$, the bottom-right block of $A_L^{-1}$. This is known as the surface green function, denoted as $g_s$. By utilizing the ideal lead assumption that removing one layer of the lead will not change $g_s$, we obtain a self-consistent form as:
\begin{align}
    g_s^{-1} = [A_{l}-A_{l,l-1}g_sA_{l-1,l}^\dagger],
    \label{sgf}
\end{align}
where $A_{l,l-1}$ is the block in $[EI-H]$ for the coupling between layer $l$ and layer $l-1$. We implemented the Lopez-Sancho algorithm \citep{sancho1985highly}, as illustrated in Algorithm, to accelerate the convergence speed. We have also implemented a modern method based on the generalized eigenvalue problem \citep{wang2008quantum} as an alternative.


\subsection{Computation of the Self-Consistent Electrostatic Potential}
\label{appendix-poisson}
Denote the charge densities in the equilibrium and non-equilibrium states as $\rho_0$ and $\rho$, and the potential fields from the original neutral and redistributed charges as $V_0$ and $V$. The equilibrium and non-equilibrium Hamiltonian can be expressed as $H_0=T+V_0$, $H_{neq}=T+V$, where $T$ is the kinetic energy. Poisson's equation relates potentials to the corresponding charge densities:

\begin{align}
\begin{cases}
\nabla\cdot\epsilon(r)\nabla V(r)=-\rho(r),\\ \nabla\cdot\epsilon(r)\nabla V_0(r)=-\rho_0(r).
\end{cases}
\end{align}

Therefore we have
$\nabla\cdot\epsilon(r)\nabla [\Delta V(r)]=-[\rho(r)-\rho_0(r)]$,
where $\Delta V=V-V_0$ is used to correct the Hamiltonian by $H_{neq}=H_0+\Delta V$. The updated $H_{neq}$ will again be used to update $\Delta V$. Hence a self-consistent iteration is constructed:
\begin{align}
\begin{cases}
    \nabla\cdot\epsilon(r)\nabla [\Delta V(r)]&=-[\rho(r;\Delta V)-\rho_0(r)], \\
    \Delta V(r)|_{\{z_L,z_R\}}&=\{V_L, V_R\}.
\end{cases}
\end{align}
Charge densities are necessary inputs for the above equation. Denote potentials in left and right electrodes as $u_l$ and $u_r$ (assume $u_l<u_r$), then the charge density $\rho(r)=-\frac{i}{2\pi}\int_{-\infty}^{+\infty}dE{G(E)}$, which can be decomposed into equilibrium and non-equilibrium terms:
\begin{align}
     \rho(r)&=\rho_{eq}(r)+\rho_{neq}(r)\\
     &=\frac{1}{\pi}Im\left[\int_{-\infty}^{u_l}dEG_D(E)\right]+\frac{1}{2\pi}\int_{u_l}^{u_r}dEG_D(E).
     \label{A10}
\end{align}
The first integration up to infinity can be computed efficiently using contour integration with the residue theorem. It is achieved by expanding the Fermi-Dirac function \citep{ozaki2007continued,areshkin2010electron}. On the other hand, the non-equilibrium charge density $\rho_{neq}$ is computed directly by numerical integration. The density of neutral charges $\rho_0$ can be computed by setting $u_l=u_r=0$.

\subsection{Expressions of Transport Properties}
\label{appendix-transport-expressions}
With the NEGF theory, electronic transport properties can be derived, such as the transmission probability ($T(E)$), the density of states ($DOS$), the electronic current ($I$), the equilibrium and non-equilibrium electronic densities ($\rho_{eq}$ and $\rho_{neq}$), etc. Here we list some of the expressions:
\begin{align}
    T(E)&=Trace[\Gamma_L(E)G_D(E)\Gamma_R(E)G_D^\dagger(E)],\\
    DOS(E)&=-\frac{1}{\pi}Trace[Im(G_D(E))],\\
    I&=\frac{2e}{\bar{h}}\int_{-\infty}^{+\infty}\frac{dE}{2\pi}T(E)[f(E-u_l)-f(E-u_r)],\label{eq_I}\\
    \rho(r)&=\frac{1}{\pi}Im\left[\int_{-\infty}^{u_l}dEG_D(E)\right]+\frac{1}{2\pi}\int_{u_l}^{u_r}dEG_D(E).
\end{align}
For Equation (\ref{eq_I}), the integral range of the current is decided by the subtraction of the Fermi-Dirac function, which is a little wider than $(u_l, u_r)$.

\section{Additional Details on Experimental Setup}
\label{appendix:repro}
The experiments are run on an Intel(R) Xeon(R) CPU E5-2650 v4 @ 2.20GHz CPU, and an NVIDIA Tesla P40 GPU. We implemented our method in PyTorch 1.9.1. We validated the correctness of our simulation results by comparing with ASE of version 3.22.0.

In the experiments, we set the learning rate of the Adam optimizer as 0.001, and the batch size as 64. Bayesian optimization is implemented based on \citet{fernando}, and the genetic algorithm is implemented based on \citet{ryan}. The bounds of the optimization variables for the black-box optimizers are ($\theta_0-0.3$, $\theta_0+0.3$), where $\theta_0$ is the initial value, namely the original 5-2 nano-junction TB Hamiltonian for the transmission curve fitting experiment, and undoped SKTB parameters for the device doping optimization experiment. The hyper-parameters of the genetic algorithm are listed in \tablename~\ref{tab:hyperparam_genetic}, and the hyper-parameters of the Bayesian Optimization algorithm are listed in \tablename~\ref{tab:hyperparam_bo}.

\begin{table}[h]
\caption{The hyper-parameters of the genetic algorithm}
\label{tab:hyperparam_genetic}
\begin{minipage}[t]{0.8\linewidth}
\begin{ruledtabular}
\begin{tabular}{lr}
\textrm{Parameter} & \textrm{Value} \\
\colrule
max\_num\_iteration & None \\
population\_size & 20 \\
mutation\_probability & 0.1 \\
elit\_ratio & 0.01 \\
crossover\_probability & 0.5 \\
parents\_portion & 0.3 \\
crossover\_type & uniform \\
max\_iteration\_without\_improv & None \\
\end{tabular}
\end{ruledtabular}
\end{minipage}
\end{table}

\begin{table}[h]
\caption{The hyper-parameters of the Bayesian optimization algorithm}
\label{tab:hyperparam_bo}
\begin{minipage}[t]{0.8\linewidth}
\begin{ruledtabular}
\begin{tabular}{lr}
\textrm{Parameter} & \textrm{Value} \\
\colrule
random\_state & 3 \\
verbose & 2 \\
kind & ucb \\
kappa & 2.5 \\
xi & 0.0 \\
\end{tabular}
\end{ruledtabular}
\end{minipage}
\end{table}

We have provided our source code in the supplementary materials for cross-checking. The code will also be released and maintained as an open-source repository in the future.


\bibliography{references}

\end{document}